\documentclass{emulateapj}
\usepackage{amsmath}

\def\H2{{{\rm H}_2}}
\def\KS{$\Sigma_{\rm SFR}-\Sigma_{\rm gas}$}
\def\MKS{$\Sigma_{\rm SFR}-\Sigma_{\rm H_2}$}
\def\nc{n_{\rm c}}
\def\CO{{\rm CO}}
\def\UMW{U_{\rm MW}}
\def\SFR{ {\mathrm SFR} }

\shorttitle{How Universal is the \MKS{} Relation?}
\shortauthors{Feldmann, Gnedin, \& Kravtsov}

\begin{document}

%=================================================
\title{How Universal is the \MKS{} Relation?}
%=================================================

\author{R. Feldmann\altaffilmark{1,2}, N. Y. Gnedin\altaffilmark{1,2,3}, and A. V. Kravtsov\altaffilmark{2,3,4}}
\altaffiltext{1}{Center for Particle Astrophysics, 
Fermi National Accelerator Laboratory, Batavia, IL 60510, USA; feldmann@fnal.gov}
\altaffiltext{2}{Kavli Institute for Cosmological Physics, The University of Chicago, Chicago, IL 60637, USA} 
\altaffiltext{3}{Department of Astronomy \& Astrophysics, The
  University of Chicago, Chicago, IL 60637, USA} 
\altaffiltext{4}{Enrico Fermi Institute, The University of Chicago,
Chicago, IL 60637, USA}

\begin{abstract}
It is a well-established empirical fact that the surface density of the star formation rate, $\Sigma_{\rm SFR}$, strongly correlates with the surface density of molecular hydrogen, $\Sigma_\H2$, at least when averaged over large ($\sim{}$kpc) scales. Much less is known, however,  if (and how) the \MKS{} relation depends on environmental parameters, such as the metallicity or the UV radiation field in the interstellar medium (ISM). Furthermore, observations indicate that the scatter in the  \MKS{} relation increases rapidly with decreasing averaging scale. How the scale-dependent scatter is generated and how one recovers a tight $\sim$ kpc scale \MKS{} relation in the first place is still largely debated.
Here, these questions are explored with hydrodynamical simulations that follow the formation and destruction of $\H2$, include radiative transfer of UV radiation, and resolve the ISM on $\sim{}60$ pc scales.
We find that within the considered range of $\H2$ surface densities (10-100 $M_\odot$ pc$^{-2}$), the \MKS{} relation is steeper in environments of low metallicity and/or high radiation fields (compared to the Galaxy), that the star formation rate (SFR) at a given $\H2$ surface density is larger, and the scatter is increased. Deviations from a ``universal''  \MKS{} relation should be particularly relevant for high-redshift galaxies or for low-metallicity dwarfs at $z\sim{}0$.
We also find that the use of time-averaged SFRs produces a large, \emph{scale-dependent} scatter in the \MKS{} relation. 
Given the plethora of observational data expected from upcoming surveys such as ALMA, the scale-scatter relation may indeed become a valuable tool for determining the physical mechanisms connecting star formation and $\H2$ formation.
\end{abstract}


%----------------------
\section{Introduction}
\label{sec:intro}
%----------------------

In a seminal paper, \cite{1959ApJ...129..243S} constructed a closed-box model of gas consumption and star formation that relies on the basic assumption of a polynomial relationship between (total) gas surface density $\Sigma_{\rm gas}$ and star formation rate surface density $\Sigma_{\rm SFR}$. This model was able to satisfy simultaneously a number of observational constraints, such as the initial luminosity function of main-sequence stars, the luminosity function of white dwarfs, or the relatively constant surface density of atomic hydrogen (HI). While the first studies focussed on the relation between neutral hydrogen and SFR \citep{1969AJ.....74...47S, 1971ApJ...163..431H}, the combination of measurements of H$\alpha$, HI and CO emission lines allowed for a direct test of the Schmidt relation, \KS{}, and a precise measurement of its exponent  \citep{1989ApJ...344..685K, 1998ApJ...498..541K}. Initially, it was assumed that  $\Sigma_{\rm gas}$ would determine $\Sigma_{\rm SFR}$ (e.g., via gravitational collapse). However, measurements of azimuthally averaged gas and SFR profiles showed that SFRs correlate better with the \emph{molecular} hydrogen ($\H2$) component than with the \emph{total} gas density \citep{2002ApJ...569..157W, 2008AJ....136.2846B}. In fact, recent observational and theoretical works demonstrate that the steepening of the \KS{} relation at low gas surface densities coincides with the transition of atomic to molecular hydrogen \citep{2008ApJ...680.1083R, 2008AJ....136.2846B, 2009ApJ...699..850K, 2010arXiv1004.0003G}. The shape of the \KS{} relation is also predicted to evolve strongly with redshift due to the 
build-up of metallicity in the interstellar medium (ISM) over cosmic history and the importance of dust in the formation of $\H2$ and its shielding from Lyman-Werner radiation \citep{2009ApJ...693..216K, 2010ApJ...714..287G}. In contrast, the \MKS{} relation is often assumed to evolve little and be relatively insensitive to changes in metallicity and interstellar radiation field, although this has not yet been confirmed observationally. 
The assumption on which this  ``universality'' is based is that the efficiency with which clouds of molecular hydrogen convert their $\H2$ into stars is not a strong function of the average ISM metallicity or the interstellar radiation field, at least under conditions typical for spiral galaxies \citep{2007ApJ...654..304K}. The scale at which this conversion takes place is the scale of (giant) molecular clouds, i.e., 100 pc or less. However, there are a couple of complications. First, there is growing observational evidence suggesting that the scatter in the \MKS{} relation increases if one goes to smaller and smaller scales (see, e.g., \citealt{2010arXiv1009.1971O, 2010arXiv1009.1651S}). Taken at face value, this seems to contradict a tight small-scale coupling between molecular hydrogen surface density and star formation. Second, the \MKS{} relation is typically measured on $\sim{}$kpc scales and the spatial averaging may lead to changes in the slope, intercept and scatter compared with those on small scales \citep{2003ApJ...590L...1K}. Third, observationally determined SFRs are time-averaged over the effective lifetime of the specific star formation tracer and may thus differ from instantaneous SFRs.

A straightforward observational check of the universality of the \MKS{} relation is difficult,  first and foremost because the direct detection of molecular hydrogen is challenging. Tracer molecules such as $\CO$ or HCN are typically used instead to infer the $\H2$ column density (e.g., \citealt{2007RPPh...70.1099O}). Mapping the line intensity of tracer molecules to the $\H2$ column density is obfuscated by the fact that the dependence of the conversion factor on ISM properties, e.g., metallicity or interstellar radiation field, is not well understood \citep{2010arXiv1003.1340G}. In addition, radiative transfer effects need to be carefully modeled (e.g., \citealt{2008ApJ...684..996N, 2010arXiv1005.3020N}).

\begin{table}
\begin{center}
\caption{Overview of the simulations discussed in this work.\label{tab:sims}}
\renewcommand{\arraystretch}{1.25}
\begin{tabular}{ccccc}
\tableline
\tableline
$\nc$ (cm$^{-3}$) & $Z/Z_\odot$ & $\UMW$ & Resolution (pc) & No. \\ \tableline
& & & &\\[-0.32cm]
50 & 0.1, 0.3, 1 & 0.1, 1, 10, 100 & 65 & 12 \\ 
$10^3$, $10^6$ & 1 & 0.1 & 65 & 2 \\ 
$10^3$, $10^6$ & 0.1 & 100 & 65 & 2 \\
50 & 1 & 1 & 32 & 1 \\
50 & 1 & 1 & 125 & 1 \\ \tableline
\end{tabular}
\tablecomments{The columns denote: (1) the minimum density of molecular clouds that form stars $\nc$, see eqution (\ref{eq:tau}), (2) the metallicity, (3) the radiation field,  (4) the spatial resolution, and (5) the number of the simulations discussed in this work.}
\end{center}
\end{table}

Numerical simulations offer a different route to studying the \MKS{} relation. What are their requirements? First, the numerical code needs to follow self-consistently the formation and destruction of $\H2$. This implies a resolution of 100 pc or better, an implementation of cooling down to a few tens of Kelvin and, also, radiative transfer of the Lyman-Werner bands (at least in some approximate form), in order to correctly capture the impact of the interstellar radiation field on the $\H2$ dissociation. 

Secondly, the code needs a recipe for star formation. The accumulating evidence in favor of a universal initial stellar mass function \citep{2010arXiv1001.2965B} indicates that star formation on small scales, i.e., within star-forming clumps and cores within molecular clouds, is largely decoupled from the ISM properties on larger scales. In particular, observations show that the average star formation efficiency per free-fall time is $\sim{}0.005-0.01$, independent of scale, once the densities of molecular clouds are reached (e.g., \citealt{2007ApJ...654..304K}, but see also \citealt{2010arXiv1007.3270M, FeldmannM}). A natural approach is therefore to couple the star formation on scales of individual molecular clouds directly to the density of molecular hydrogen, assuming that the formation and destruction of $\H2$ can be modeled reliably. 

Consequently, the approach we use in our simulations is to relate the SFR to the $\H2$ density \emph{on small scales} ($\sim{}60$ pc) via the following equation \citep{2009ApJ...697...55G}:
\begin{equation}
\label{eq:SFR}
\dot{\rho}_* = \epsilon_{\rm SFR} \frac{\rho_{\rm H}}{\tau_{\rm SFR}}f_\H2{} \,e^{ \sigma{}X  - \frac{1}{2}\sigma^2 },
\end{equation}
where $\dot\rho_*$ is the \emph{instantaneous} SFR density, $\rho_{\rm H}$ is the hydrogen mass density, $f_\H2{}$ is the $\H2$ fraction, $\epsilon_\mathrm{SFR}$ and $\tau_\mathrm{SFR}$ denote the star formation efficiency and the star formation time-scale (see equation \ref{eq:tau} below), respectively. 
The exponential factor models intrinsic scatter with width $\sigma$ (we discuss intrinsic scatter in \S \ref{sect:scale}, otherwise we assume $\sigma=0$). $X$ is a Gaussian random variable with mean 0 and variance 1. The factor $e^{- \frac{1}{2}\sigma^2}$ ensures that the same amount of gas is converted into stars at a given $\H2$ density \emph{independent} of the scatter width $\sigma$. This implies, however, that star formation efficiencies derived from the mean (or median) relation in the $\log\dot{\rho}_* - \log\rho_\H2$ plane, i.e., for $X=0$, differ from the parameter $\epsilon_{\rm SFR}$ by the factor $e^{- \frac{1}{2}\sigma^2}$. This effect should be considered when, e.g., star formation efficiencies of scatter-free numerical simulations are calibrated against observations on sub-kpc scales.

On which grounds would we actually expect to see any dependence of slope, intercept and scatter of the \MKS{} relation on  environmental parameters such as metallicity or interstellar radiation field? The $\H2$ abundance is strongly affected by the amount of dust shielding from the UV radiation and, consequently, a lower metallically and/or larger radiation field will increase the required density for $\H2$ (and consequently stars) to form. We will show that a non-linear relation between $n_\H2$ and the SFR on small scales can have a significant impact on the slope, intercept, and scatter of the \MKS{} relation measured on large ($\sim{}$ kpc) scales. Another important, and so far often neglected quantity, is the scatter in the \MKS{} relation. While some scatter may be due to observational measurement uncertainties, it is clear that any environmental dependence of the \MKS{} relation will translate into a galaxy-to-galaxy variation and, in combined data sets, to scatter. Furthermore, the observed \MKS{} relation is measured on large scales (spatial averaging) using time-averaged SFRs. The averaging may induce a scale-dependence of the scatter.

The layout of the paper is as follows. In \S \ref{sec:sims} we briefly describe the setup of our numerical experiments. We then show in \S \ref{sect:UVZ}  the predicted dependence of the slope, intercept, and scatter of the \MKS{} relation on metallicity and interstellar radiation field. The scale dependence of the scatter and the propagation of intrinsic scatter from 100 pc to kpc scales is studied in \S \ref{sect:scale}. We discuss our findings in \S \ref{sect:Discussion} and conclude in \S \ref{sect:Conclusion}.

%---------------------
\section{Simulations}
\label{sec:sims}
%---------------------

A detailed description of the set of performed simulations can be found in \cite{2010arXiv1004.0003G}. All simulations are run with the Eulerian hydrodynamics + $N$-body code ART \citep{1997ApJS..111...73K, 2002ApJ...571..563K}, which uses an adaptive mesh refinement (AMR) technique to achieve high spatial resolution in the regions of interest (here: regions of high baryonic density). First, we ran an initial cosmological, hydrodynamical simulation down to $z=4$. This simulation follows a Lagrangian region that encloses five virial radii of a typical $L_*$ galaxy (halo mass $\sim{}10^{12}$ $M_\odot$ at $z=0$) within a box of 6 comoving Mpc/$h$. The mass of dark-matter particles in the high-resolution Lagrangian patch is $1.3\times{}10^6 M_\odot$ and the spatial resolution is $65$ pc at $z=3$ in physical coordinates. We adopt the following cosmological parameters: $\Omega_{\rm matter}=0.3$, $\Omega_\Lambda=0.7$, $h=0.7$, $\Omega_{\rm baryon}=0.043$, and $\sigma_8=0.9$. This initial, fully self-consistent simulation is consequently continued for additional $\sim{}600$ Myr before it is analyzed, but now with metallicities and UV fields  fixed to a specific, spatially uniform value. At this time, the mass of the simulated halo is $\approx{}4.2\times{}10^{11}$ $M_\odot$. We have run a grid of simulations with three different metallicities $Z=0.1, 0.3, 1.0$ (in units of $Z_\odot=0.02$) and four different values of the interstellar radiation field $\UMW=0.1,1,10,100$. The parameter $\UMW=J/J_{\rm MW}$ specifies the strength of the interstellar radiation field in units of the radiation field of the Milky Way at 1000\AA: $J_{\rm MW}=10^6$ photons cm$^{-2}$ s$^{-1}$ ster$^{-1}$ eV$^{-1}$ \citep{1978ApJS...36..595D, 1983A&A...128..212M}. We continued one of our simulations ($Z/Z_\odot=1$, $\UMW=1$) for additional 400 Myr and found no significant changes in the \MKS{} relation. This indicates that the predictions of our simulations should also hold for redshifts $z\lesssim{}3$, at least unless/until ISM properties change radically. In total, we ran a set of 18 simulations (including 2 runs for a resolution study) in order to explore the effect of varying metallicity, radiation field and density threshold on the \MKS{} relation (see Table \ref{tab:sims}).

The molecular hydrogen fraction $f_\H2{}$ is computed self-consistently, including a chemical network comprised of 6 species and radiative transfer of the UV continuum and the Lyman-Werner bands \citep{2010arXiv1004.0003G}. If the average density in a simulation cell is smaller than the density typical for molecular clouds, we have to revert to a `subgrid' interpretation of the $\H2$ fraction. In this case, we assume that the fraction $f_\H2$ corresponds to the (mass) fraction of hydrogen in individual (unresolved) molecular clouds. 

\begin{figure*}
\begin{tabular}{cc}
\includegraphics[width=85mm]{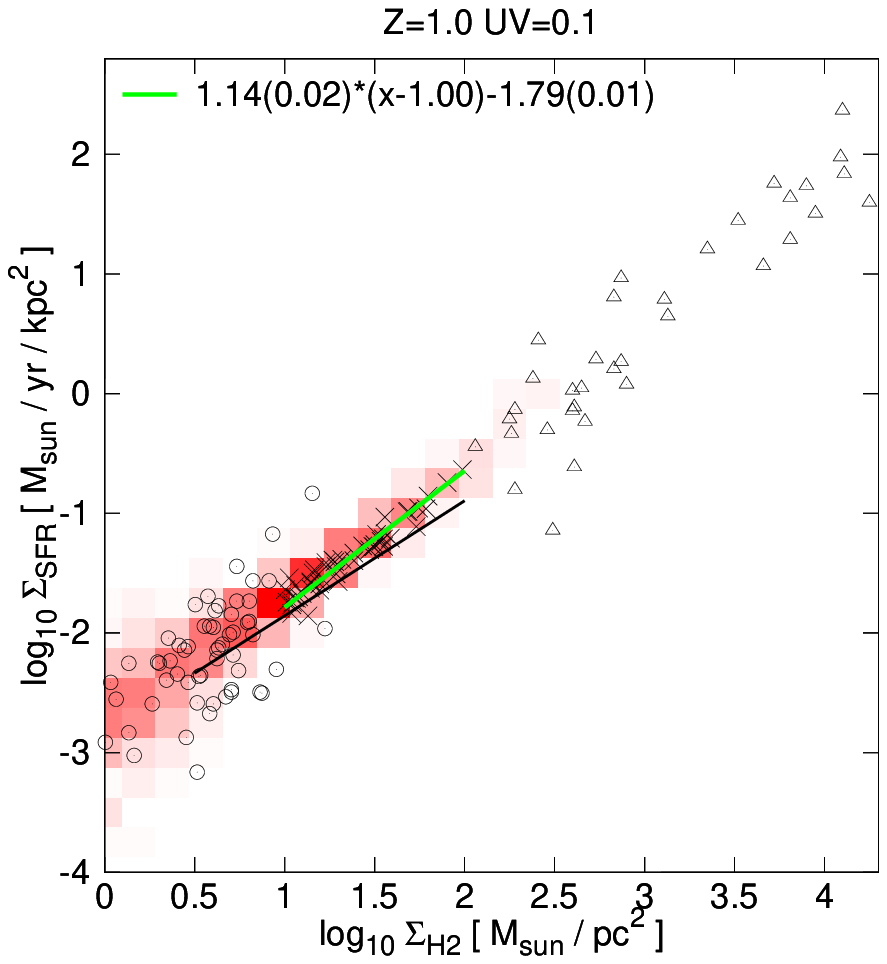} &
\includegraphics[width=85mm]{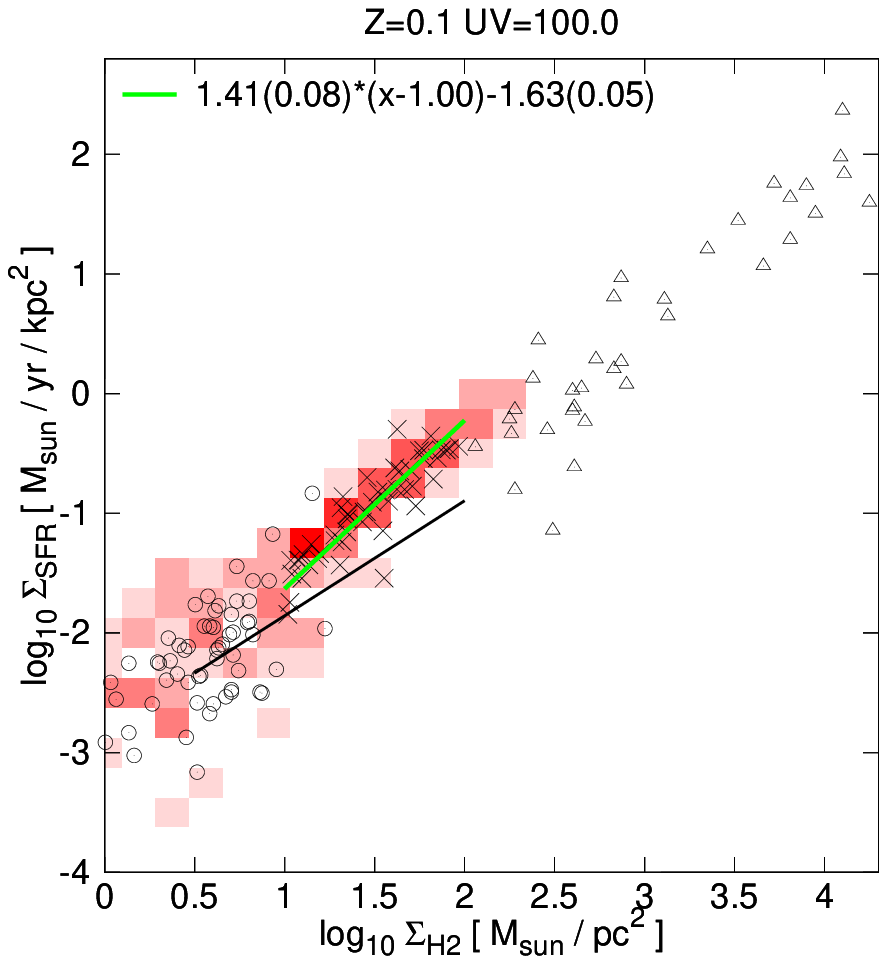}
%\plottwo{f1a}{f1b}
\end{tabular}
\caption{\MKS{} relation on the kpc scale. 
Left panel:  $Z=Z_\odot$, $\UMW=0.1$. Right panel: $Z=0.1\,Z_\odot$, $\UMW=100$.
Stellar and $\H2$ masses are measured within cubical cells of $l=1$ kpc box length. Surface densities are estimated by dividing each mass by $l^2$. 
SFRs are averaged over 20 Myr.
The simulation results are shown by the red shaded region (two-dimensional histogram of all simulation cells) and by crosses (a random sample of
50 simulation cells with surface density in the range $10<\Sigma_\H2 / M_\odot \mathrm{pc}^{-2}<100$). 
The solid green line is the result of a bisector regression of all kpc-sized simulation cells with $10<\Sigma_\H2 / M_\odot \mathrm{pc}^{-2}<100$ and $0.01<\mathrm{SFR} / M_\odot \mathrm{yr}^{-1} \mathrm{kpc}^{-2}<1$. The regression parameters, slope and intercept, are shown on the top left. Also shown (in parentheses) are the regression errors, estimated via bootstrapping.
The black circles and triangles correspond to the normal spiral and star bursting sample, respectively, of \cite{1998ApJ...498..541K}. The solid black line is the average \MKS{} relation found in \cite{2008AJ....136.2846B}. The \MKS{} relation in the right panel has a steeper slope, a higher normalization, and a larger scatter than in the left panel.
\label{fig:KSrelation}}
\end{figure*}

Consequently, the star formation timescale is given by the minimum of (1) the free-fall time corresponding to the average density in the cell and (2) the free-fall time corresponding to the minimum density of molecular clouds that form stars $\nc$, i.e.,
\begin{equation} 
\label{eq:tau}
\tau_\mathrm{SFR}=\min[\tau_\mathrm{ff}(n_{\rm H}), \tau_\mathrm{ff}(\nc)].
\end{equation}

% figure 1

We stress that for densities smaller than $\nc$ the relation between SFR and $\H2$ abundance is linear, while it becomes non-linear for larger densities, because
$\tau_\mathrm{ff}(n_{\rm H})\propto{}n_{\rm H}^{-1/2}$. A non-linear steepening of the \MKS{} relation at $\Sigma_\H2>100 M_\odot$ pc$^{-2}$  is motivated by theoretical studies (e.g., \citealt{2009ApJ...699..850K}), but
not yet confirmed by observations. We therefore explore the case in which $\nc=50$ cm$^{-3}$, i.e., close to the typical average density of molecular clouds ($\sim{}100-200$ cm$^{-3}$), but also discuss the possibility of much larger thresholds such as $\nc=10^3$ cm$^{-3}$ and $\nc=10^6$ cm$^{-3}$.  Since our simulations do not capture densities of $\gtrsim{}10^5$ cm$^{-3}$, a threshold above this value effectively corresponds to a fully linear SFR - $\H2$ relation on small scales.

Instantaneous SFRs are computed directly using equations (\ref{eq:SFR}) and (\ref{eq:tau}). Our simulations use $\epsilon_\mathrm{SFR}=0.005$. This value, which is consistent with small-scale observations \citep{2007ApJ...654..304K}, ensures that the normalization of the \MKS{} relation on kpc scales is similar in simulations and observations. Time-averaged SFRs (over time $T$) are calculated by counting the number of stars in a cube of given scale with ages below $T$. Unless otherwise noted, we use $T=20$ Myr, but we have explicitly checked that our results do not change significantly if larger averaging times are used (up to $T=200$ Myr). SFR estimates based on observations of UV luminosities in the wavelength range 1250-2800 \AA{} correspond to an averaging time of $\sim$ 100 Myr, which based on nebular emission lines, such as $H\alpha$, typically correspond to $T\sim{}10$ Myr, and estimates based on the FIR continuum (e.g., 24 $\mu{}$m) correspond to a range ($\sim{}$10 - 100 Myr) of averaging time scales; see, e.g., \cite{1998ARA&A..36..189K}.

The slope and intercept of the \MKS{} relation are obtained with a bisector regression in log-log space \citep{1990ApJ...364..104I}. Although the use of bisector regression cannot be rigorously justified in general (see \citealt{2007ApJ...665.1489K, 2010arXiv1008.4686H}), the bisector method is sufficient for our purposes as we perform regression on tightly correlated data without error bars. We estimate scatter about the best-fit relation, as the root mean square of log10 of the spatially averaged star formation rate density relative to its value on the regression line with the same density, see also equation (\ref{eq:sigSFR}). We estimate errors for the slope, intercept, and scatter using the standard bootstrap method \citep{Efron1979} with a sample size of 200.

\section{Results}

\subsection{Dependence on metallicity and UV field}
\label{sect:UVZ}
 
\begin{figure*}
\begin{tabular}{c}
\includegraphics[width=175mm]{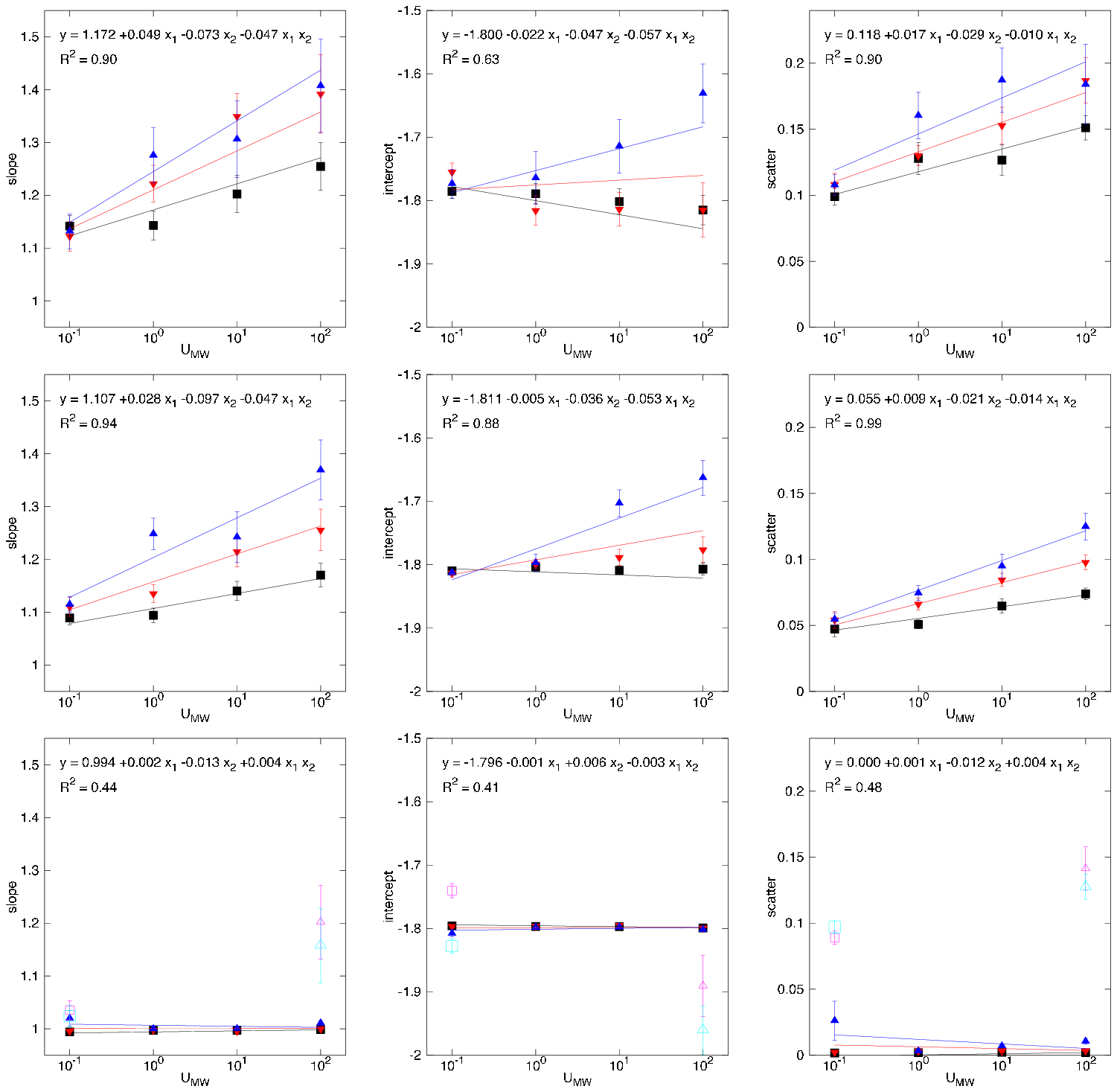}
%\epsscale{0.85}
%\plotone{f2}
\end{tabular}
\caption{\small{}Dependence of slope, intercept, and scatter on metallicity and interstellar radiation field. The \emph{first} row shows (from left to right) the slope, the intercept, and the scatter
of the \MKS{} relation (averaged over kpc scales) as a function of the radiation field, $\UMW$ ($x$-axis), and for different metallicities ($Z/Z_\odot=1$ (black squares), $0.3$ (red downward-pointing triangle), and $0.1$ (blue upward-pointing triangle)). SFRs are averaged over 20 Myr and the minimum cloud density is $\nc=50$ cm$^{-3}$. 
Slope, intercept and scatter are computed from a bisector regression of the \MKS{} relation as described in the caption of Fig.~\ref{fig:KSrelation}.
To highlight the trends with $Z$ and $\UMW$, we also performed a two-parametric regression of slope, intercept, and scatter as a function of $Z$ and $\UMW$ (regression equation and parameters and the square of the correlation coefficient are shown at the top of each panel; $x_1=\log_{10}\UMW$, $x_2=\log_{10}Z$).  The black, red, and blue solid lines (from bottom to top) show the results of the bi-parametric regression for the choices $Z/Z_\odot=1$, $0.3$, and $0.1$, respectively.
The \emph{middle} row shows the same quantities as the top row, but for instantaneous SFRs. The \emph{bottom} row shows again the same quantities, but for a larger threshold density $\nc$. Specifically, the filled symbols and lines use instantaneous SFRs and $\nc=1000$ cm$^{-3}$, while the empty symbols
use time-averaged SFRs and $\nc=1000$ cm$^{-3}$ (small magenta symbols) and $\nc=10^6$ cm$^{-3}$ (large cyan symbols), respectively. We note that whenever $\nc>50$ cm$^{-3}$ the star formation efficiencies are reduced by $\sqrt{\nc/50}$ (see equations (\ref{eq:SFR}) and (\ref{eq:tau})) in order to ensure the correct normalization of the \MKS{} relation.
\label{fig:regfit} }
\end{figure*}
  
In Fig. \ref{fig:KSrelation} we plot and compare the \MKS{} relation for (1) solar metallicity and $\UMW=0.1$, and (2) $Z/Z_\odot=0.1$ and $\UMW=100$. Measured over the range 
$10<\Sigma_\H2 / M_\odot \mathrm{pc}^{-2}<100$, the slope of the relation in case (1) is $\sim{}1.14\pm{}0.02$, the SFR at a surface density $\Sigma_\H2=10$ $M_\odot$ pc$^{-2}$ is 0.016 $M_\odot$ yr$^{-1}$ kpc$^{-2}$ and the scatter of
$\log_{10}{\rm SFR}$ around the best fit is 0.10 dex. The slope is slightly steeper than that derived from CO measurements ($\sim{}0.96\pm{}0.07$; \citealt{2008AJ....136.2846B}). 
As anticipated, the choice $\epsilon_\mathrm{SFR}=0.005$ leads to a normalization of the simulated \MKS{} relation that is close to what is found in observations, once observational data are mapped to the same initial stellar mass function (IMF). In case (2), the slope is significantly steeper $\sim{}1.4$, the SFR at a surface density $\Sigma_\H2=10$ $M_\odot$ pc$^{-2}$ higher (0.023 $M_\odot$ yr$^{-1}$ kpc$^{-2}$) and the scatter is larger (0.18 dex).

%\epsscale{0.8}
\begin{figure}
\begin{tabular}{c}
\includegraphics[width=92mm]{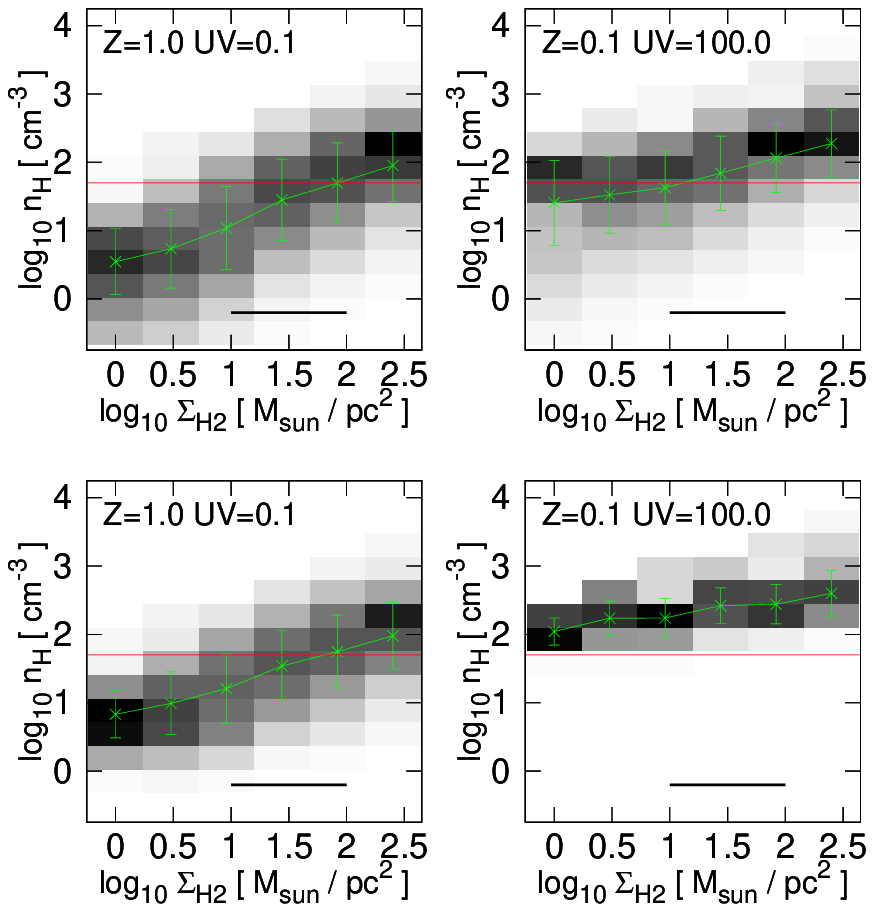} 
%\plotone{f3}
\end{tabular}
\caption{
Distribution of neutral hydrogen mass (top row) and $\H2$ mass (bottom row) as a function the small-scale ($\sim{}60$ pc) hydrogen density $n_\mathrm{H}$ (ordinate) for several large-scale (1 kpc) $\H2$ surface densities (abscissa). Each vertical ($\H2$ surface density) bin is normalized to the total mass of neutral hydrogen (top row) or $\H2$ (bottom row) in the bin. The grade of shading indicates the mass fraction (on a linear scale). A black pixel contains 100\% of 
the neutral hydrogen or $\H2$, respectively, in the given $\H2$ surface density bin; white corresponds to 0\%. The left column corresponds to a galaxy with
high metallicity and low radiation field ($Z=Z_\odot$, $\UMW=0.1$), while the right column is for a low-metallicity, high radiation field galaxy ($Z=0.1\,Z_\odot$, $\UMW=100$). Mean and standard deviation of the density distributions of neutral hydrogen (top row) or $\H2$ (bottom row) are indicated by green crosses and error bars. The thin, red horizontal line indicates the threshold $n_c=50$ cm$^{-3}$. The thick, black horizontal line near the bottom of each panel shows the range over which slope, intercept, and scatter of the \MKS{} relation are computed, see Fig. \ref{fig:regfit}.
\label{fig:SH2NB}}
\end{figure}

In Fig. \ref{fig:regfit} we plot the slope, intercept, and scatter of the \MKS{} relation, spatially averaged over 1 kpc, for a grid of environmental parameters.
The figure shows that the slope, intercept and scatter are systematically changing as a function of $Z$ and $\UMW$. A bi-parametric regression (using $Z$ and $\UMW$ as independent variables) captures the change in slope and scatter very well. The regression parameters are given in the legend of the figure.
A word on the terminology. We refer to the scale at which equations (\ref{eq:SFR}) and (\ref{eq:tau}) are applied as ``small scales'' ($\sim{}60$ pc in our simulations).
By contrast, we refer to the scales on which the slope and intercept of the \MKS{} are measured as ``large scales'' ($\sim$ kpc in this study).
The spatial density on scale $l$ is given by the amount of mass within a cube of size $l$. In order to convert from a spatial to a surface density, we multiply the spatial density by $l$. We do not use the surface density on the smallest ($\sim{}60$ pc) scales in order to avoid underestimating it.

Fig. \ref{fig:regfit} shows that the use of time-averaged SFRs introduces the \emph{dominant} amount of scatter in the \MKS{} relation on large scales. Specifically,
as the first two panels in the rightmost column demonstrate, the scatter in the relation is significantly larger ($\sim{}0.1-0.2$ dex) if SFRs are time averaged, compared
with the case that instantaneous SFRs are used ($\sim{}0.05-0.12$ dex). Time averaging creates scatter because $\H2$ surface densities are measured instantaneously while the SFRs are averaged over some past time interval.

However, the use of time-averaged SFRs is not the only source of scatter. The important point to realize is that equation (\ref{eq:SFR}) depends both on the $\H2$ density ($\rho_\H2=\rho_{\rm H}f_\H2$) \emph{and} the hydrogen density $\rho_{\rm H}$ (via $\tau_{\rm SFR}$). Hence, on small scales, a scatter in the hydrogen density at \emph{fixed} $\H2$ density translates into a scatter of SFR at fixed $\H2$ surface density. The value of the threshold $\nc$ affects this type of scatter in a crucial way. If $\nc$ is very large (much larger than the peak in the mass-weighted distribution function of molecular hydrogen) then the SFR does not depend explicitly on $\rho_{\rm H}$ (since $\tau_{\rm SFR}=\tau_{\rm ff}(\nc)$) and, consequently, no scatter is generated. Similarly, for hydrogen densities above a certain limit, let us call it $n_{\rm fm}$\footnote{$n_{\rm fm}\sim{}300$ cm$^{-3}$ for $\UMW=100$, $Z/Z_\odot=0.1$ and $n_{\rm fm}\sim{}10$ cm$^{-3}$ for $\UMW=0.1$, $Z/Z_\odot=1$.}, the gas is fully molecular and, hence, $\rho_{\rm H}$ and $\rho_\H2$ are 1:1 related (see, e.g., \citealt{2009ApJ...697...55G}). If $n_{\rm H}>n_{\rm fm}$, no scatter is produced on the level of a single cell, but scatter can still arise on \emph{larger} scales as cells with different properties are added. To clarify this point, let us assume that we add the SFRs and $\H2$ densities from, e.g., two cells $A$ and $B$. First, let cell $A$ have a density below $\nc$ and cell $B$ a density above $n_{\rm fm}$. Second, let us redistribute the hydrogen and $\H2$ masses such that \emph{both} cells have a density below $\nc$ (this might not be possible in all cases). Although in both cases the $\H2$ density is the same, the SFRs are higher in the first case.

The mechanism that we have just described explains the \emph{existence} of scatter, provided $\nc$ is sufficiently low (see the third column of Fig. \ref{fig:regfit}). However, we have not discussed why there is a \emph{trend} of scatter with $Z$ and $\UMW$. The origin of this trend can be understood from the bottom panels of Fig. \ref{fig:SH2NB}, where we show a histogram of the small-scale hydrogen density ($\H2$ mass-weighted) parametrized by the large-scale $\H2$ surface density. The figure shows that the fraction of $\H2$ mass that is in cells with hydrogen densities above a given threshold (in the range of $\sim{}10-100$ cm$^{-3}$) increases with decreasing $Z$ and increasing UV. Hence, more of the $\H2$ mass participates in producing scatter and the overall scatter increases.

Fig. \ref{fig:regfit} also shows that there is a dependence of the intercept of the \MKS{} relation on $Z$ and $\UMW$, provided $\nc$ is sufficiently small (see the middle panel in the top and middle rows). How do we understand this result? As we just pointed out, an increase in the radiation field and/or a decrease in the metallicity shifts the peak of the mass weighted $\H2$ density distribution towards higher densities (see the bottom row of Fig. \ref{fig:SH2NB}). Specifically, the figure shows that for  $\UMW=0.1$ and $Z/Z_\odot=1$, cells with hydrogen densities in the range of $\sim{}3-100$ cm$^{-3}$ contain most of the molecular hydrogen
for the considered range of $\H2$ surface densities. Hence, only cells in the range of $\sim{}3-100$ cm$^{-3}$ contribute significantly to the SFR. On the other hand, if $\UMW=100$ and $Z/Z_\odot=0.1$, only cells with hydrogen densities in the range of $100-500$ cm$^{-3}$ contribute to star formation. Hence, a low-metallicity, high UV disk will only keep ``pockets'' of $\H2$ in high density regions, while in a high-metallicity, low UV disk $\H2$ is present even in much lower density regions. Consequently, for a given large-scale $\H2$ surface density, more of the $\H2$ sits at high densities in a low $Z$, high UV field, galaxy compared to a high $Z$, low UV field galaxy.
Furthermore, in the regime in which $n_{\rm H}>\nc$ and the hydrogen gas is (close to) fully molecular the SFRs scale as $\propto{}n_{\rm H}^{1.5}$. 
This \emph{non linearity} then amounts to a higher SFR (and hence intercept) \emph{at fixed large-scale $\H2$ surface density} in a low $Z$, high UV field, galaxy compared to a high $Z$, low UV field galaxy. In other words, the large-scale SFRs depend not only on the large-scale $\H2$ surface densities, but also on the distribution function of $n_{\rm H}$ on small scales. 

A related mechanism leads to a dependence of the slope on $Z$ and $\UMW$ (left panel in the top and middle rows of Fig.\ref{fig:regfit}). The panels in the top row of Fig. \ref{fig:SH2NB} show that the typical hydrogen densities of cells that contribute to a given $\H2$ surface density $\Sigma_\H2$ increase with $\Sigma_\H2$. This demonstrates that the density structure of disks \emph{at a given $\H2$ surface density} changes with metallicity and radiation field of the ISM. The bottom panels in Fig. \ref{fig:SH2NB} show that this trend remains (although somewhat weakened) if the hydrogen density distribution is weighted by $\H2$ mass. It also shows, that the effect of an increasing hydrogen density with large-scale $\H2$ surface density is stronger for a high $Z$, low $\UMW$ galaxy than for a low $Z$, high $\UMW$ galaxy. However, in the former case, most of the $\H2$ is in cells with densities $n_{\rm H} < \nc$ and consequently the SFR density is still proportional to the total $\H2$ density, i.e., slope 1. In the latter case, however, this increase is important. Let us see why. From (\ref{eq:SFR}) and (\ref{eq:tau}) it is clear that the surface density of the SFR is proportional to $\sum_i n_{\H2,i} n_{{\rm H},i}^{\alpha_i}$, where the sum is over all cells within the given line-of-sight cylinder and $\alpha$ is either 0 (if $n_{\rm H}<\nc$) or 0.5 (if $n_{\rm H}\geq{}\nc$). The $\H2$ surface density is proportional to $\sum_i n_{\H2,i}$. An increase in the $\H2$ surface density $\Sigma_{\H2}\rightarrow{}\gamma{}\Sigma_{\H2}$ can be achieved in several ways. If the length of the cylinder increases, then the surface density of the SFR increases proportional to $\Sigma_{\H2}$. If, however, the density structure changes (in the simplest case via $n_{\rm H}\rightarrow{}\gamma{}n_{\rm H}$), then the surface density of the SFR increases by $\gamma{}^{1+\alpha}$ (assuming the gas is fully molecular). Obviously, if $\alpha=0$ (as for $n_{\rm H}<\nc$) the predicted large-scale slope of the \MKS{} relation is, as expected, linear. However, it lies between 1 and $1+\alpha$ if there is a mix of cells with densities below and above $\nc$. In addition, if the density distribution changes in a more complicated manner with $\Sigma_{\H2}$, it is also possible to obtain ``rolling'' slopes or even large-scale slopes that are steeper than 1.5. We conclude that \emph{the slope of the \MKS{} relation on kpc scales can vary and depends on the $\H2$ density distribution as a function of the large-scale $\H2$ surface density}, see also \cite{2003ApJ...590L...1K}.

While the time averaging of the SFRs generates most of the scatter in the \MKS{} relation, the trends of slope and scatter with $Z$ and $\UMW$ are largely driven by the non-linear coupling between SFR and $\H2$ density. This can be clearly seen in the last row of Fig. \ref{fig:regfit}. If $\nc=10^6$ cm$^{-3}$, the slope of the \MKS{} relation changes only between 1.03 ($Z/Z_\odot=1$, $\UMW=0.1$) and 1.16 ($Z/Z_\odot=0.1$, $\UMW=100$) and the scatter increases only from 0.09 dex to 0.12 dex. We discuss the dependence of the scatter on ISM properties further in the next section.

If SFRs are measured instantaneously \emph{and} the small-scale relation between star formation rate density and $\H2$ density is linear (i.e., $\nc$ is large),
then the slope reduces to exactly unity, and any dependence of the intercept on metallicity or radiation field is eliminated and the scatter vanishes (at least
as long as there are no other sources of scatter, see \S\ref{sect:scale}). The reason lies in the fact that 
spatial averaging (which is a linear operation) over a linear relation between SFR and $\H2$ density on small scales results again in a linear \MKS{} relation on larger scales.

%\item \textbf{An important result of our study is that changes in the small scale star formation model (e.g., by changing $\nc$) can lead to different predictions in the \MKS{} relation on large-scales. This complements previous studies that reach a similar conclusion for the relation between SFR and total gas mass (the Kennicutt-Schmidt relation), e.g., \cite{2003ApJ...590L...1K, 2008MNRAS.383.1210S}.}

We conclude that slope, intercept, and scatter of the \MKS{} relation averaged on kpc scales can change systematically with
metallicity and radiation field. While our quantitative predictions likely depend on the assumed star formation model\footnote{For instance, as we show in Fig.~\ref{fig:regfit}, they vary with $\nc$.} and the model for $\H2$ formation and shielding the \emph{existence} of a ``non-universality'' of the large-scale \MKS{} relation is a rather generic outcome whenever there is a non-linear relation between SFR and $\H2$ density on small scales.
 
% figure 3

\subsection{Dependence on averaging scale}
\label{sect:scale}

Observational studies show that the \MKS{} relation has larger scatter on smaller scales \citep{2010A&A...510A..64V, 2010arXiv1009.1971O, 2010arXiv1008.3183D}. Specifically, recent observations of CO, H$\alpha$, and 24$\mu$m emission in M33 have been used to argue that the \MKS{} relation ``breaks down'' on a scale of $\sim{}100$ pc. It has been suggested that the drifting of newly formed star clusters or the difference in evolutionary stages of molecular clouds / star clusters could be responsible \citep{2010arXiv1009.1971O}. Given the limited range of measured gas surface densities, it is plausible that this ``break-down'' is merely a manifestation of a very large scatter that may arise from a variety of sources. In \cite{2010arXiv1009.1971O}, the studied range of surface densities is approximately $1 M_\odot{\rm pc}^{-2} < \Sigma_\H2 < 10  M_\odot{\rm pc}^{-2}$. By contrast, the average gas surface density of GMC measured on scales of a few tens of parsec in M33 is $\sim{}120 M_\odot{\rm pc}^{-2}$ \citep{2003ApJ...599..258R}, roughly similar to that found in the Milky Way \citep{1987ApJ...319..730S, 2009ApJ...699.1092H}. Hence, $\H2$ gas surface densities below $\lesssim{}10 M_\odot/{\rm pc}^2$ measured on 100 pc scales must correspond to the outskirts of GMCs, not to GMCs themselves. This by itself may be responsible for a substantial fraction of the measured scatter. We, instead, will focus on a range of 10 times larger $\H2$ surface densities, namely $10 M_\odot/{\rm pc}^2 < \Sigma_\H2 < 100  M_\odot/{\rm pc}^2$ for scales from kpc down to 100 pc. Instead of aiming at a precise quantitative comparison with observations (that are not yet available), we discuss qualitative predictions of our simulations, their limitations, and the resulting implications.

\begin{figure}
\begin{tabular}{c}
\includegraphics[width=92mm]{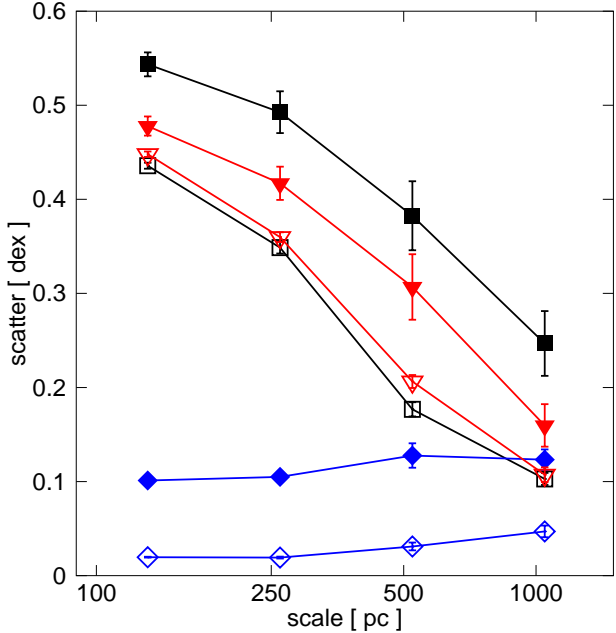} 
%\plotone{f4}
\end{tabular}
\caption{Dependence of the scatter in the \MKS{} relation on the averaging scale. The scatter in the \MKS{} relation has been derived assuming either (1) time-averaged SFRs (20 Myr) and a minimum cloud density of $\nc=50$ cm$^{-3}$ (black squares), (2) time-averaged SFRs and $\nc=10^6$ cm$^{-3}$ (red triangles), or (3) instantaneous SFRs and  $\nc=50$ cm$^{-3}$ (blue diamonds). The scatter is computed with an ordinary least squares regression of the SFR as a function of $\H2$ surface density in the range $10<\Sigma_\H2 / M_\odot \mathrm{pc}^{-2}<100$. Errors are computed via bootstrapping. Empty symbols correspond to solar metallicity and $\UMW=0.1$, while filled symbols refer to $Z=0.1\,Z_\odot$ and $\UMW=100$. Simulations with intermediate values of metallicity and UV field lie in between. \label{fig:scaleScatter}}
\end{figure}

In Fig.~\ref{fig:scaleScatter} we show the scatter of the \MKS{} relation as a function of scale from 1 kpc down to $\sim{}$100 pc. 
The first thing we notice is that the scatter due to time-averaging alone (red lines and triangles) increases with decreasing scale,
while the scatter solely due to the threshold density $\nc$ (blue lines and diamonds) remains roughly scale-independent. The origins of the different types of scatter have been
discussed in the last section. The figure shows that the scatter on all scales is primarily caused by the time averaging of the SFR. This is not an artifact of the particular SFR averaging time used. We varied the SFR averaging time scales between 20 and 200 Myr and found no substantial change in the amount of scatter, as long as low SFR outliers ($>$ 3 sigma) are excluded\footnote{If not excluded, these outliers do increase the scatter somewhat (by $\sim{}0.1$ dex) when averaging over 200 Myr instead of 20 Myr.}. 

% Comp with obs
A straightforward comparison between our predictions and observations \citep{2010A&A...510A..64V, 2010arXiv1009.1971O, 2010arXiv1009.1651S}  is difficult as none of the published observations explicitly quantify the scatter as a function of scale. However, the scatter predicted from our simulations ($\sim{}0.4-0.6$ dex on $\sim{}$100 pc, $\sim{}0.1-0.3$ on $\sim{}$kpc scales) seems to be of similar order as, e.g., the observed scatter in Fig. 4 of \cite{2010A&A...510A..64V} and the scatter in Fig. 3 of \cite{2010arXiv1009.1651S}. One should keep in mind though, that the scatter in the literature may include contributions from observational uncertainties. Also, the standard deviation in \cite{2010A&A...510A..64V} is computed on iteratively sigma-clipped data, which likely underestimates the scatter. Even more problematic is that all mentioned observational studies derive $\H2$ masses from $\CO$ measurements, which may introduce additional scatter. A detailed modeling of the H2-CO conversion factor which would be required for a fair comparison is beyond the scope of this paper.

Fig.~\ref{fig:scaleScatter} also shows that the scatter due to the SFR time averaging depends to a small extent on $\UMW$ and $Z$. This result can be rephrased in terms of a duty fraction, which we define as the fraction of time (the relevant time scale is the SFR averaging time) during which the $\H2$ density within the considered cell is close to its time-averaged value. A duty fraction of unity does not introduce scatter on small scales, as it means that within the SFR averaging time the $\H2$ content within the cell remains constant. Fig.~\ref{fig:scaleScatter} then shows that the duty fraction decreases with increasing radiation field/decreasing metallicity, hence leading to larger scatter. One interpretation of the reduced duty fraction is that stronger $\UMW$ and/or lower $Z$ reduce the life times of molecular clouds. An alternative possibility is that molecular clouds live as long as before, but molecular cloud formation is rarer.

Another potential contributor to the scatter on small scales is the velocity spread of young stellar clusters and the stars within the cluster. 
This effect is not modeled adequately in the simulation because we do not resolve individual cluster members, but rather obtain one `star particle' for each cluster that initially moves with the average velocity of the gas. On scales $\gtrsim{}100$ pc this effect plays 
only a small role presumably, as the typical distance that stars travel within 20 Myr is of the order of $\sim{}100$ pc (assuming an rms velocity of  $\sim{}5$ km s$^{-1}$). Some scatter on large scales may arise from high-velocity run-away stars \citep{1961BAN....15..265B, 1991AJ....102..333S}.

\begin{figure}
\begin{tabular}{cc}
\includegraphics[width=92mm]{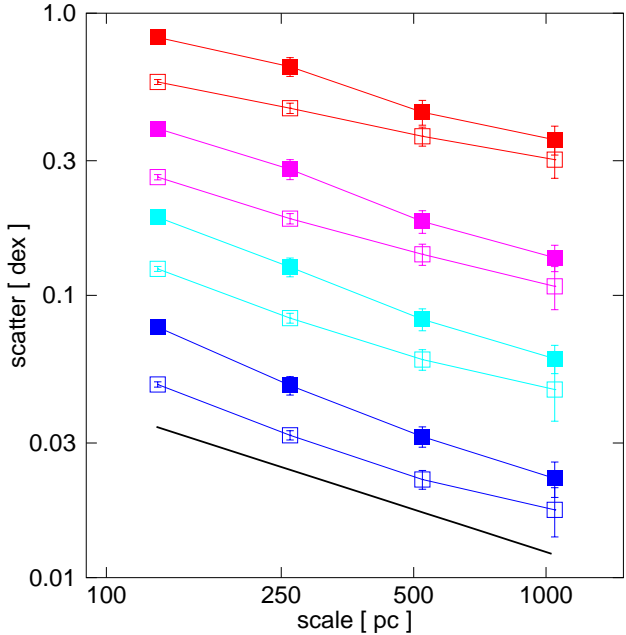} 
%\plotone{f5}
\end{tabular}
\caption{Propagation of scatter in the \MKS{} relation from small to large scales. A scatter in $\log_{10}(\mathrm{ SFR } [ M_\odot \mathrm{yr}^{-1} ])$ of (from bottom to top) 0.1 (blue lines), 0.25 (cyan lines), 0.5 (magenta lines), and 1 (red lines) is inserted at the resolution scale (65 pc) of the simulation. The black line at the bottom shows a power law: scatter $\propto{}{\rm scale}^{-0.5}$. The scatter in the  \MKS{} relation is measured in the range $10<\Sigma_\H2 / M_\odot \mathrm{pc}^{-2}<100$. Instantaneous SFRs and a minimum cloud density of $10^6$ cm$^{-3}$ are used in order to suppress other sources of scatter. Filled (empty) symbols correspond to the simulation with $Z=0.1\,Z_\odot$, $\UMW=100$ ($Z=Z_\odot$, $\UMW=0.1$).\label{fig:scaleScatter2}\vspace{1cm}}
\end{figure}

The decline of the scatter with increasing averaging scale is obviously related to the spatial averaging over a larger number of resolution elements $N_{\rm res}$. Naively, we would expect a scaling proportional to $1/\sqrt{N_{\rm res}}$, where $N_{\rm res}\propto{}l^3$ if the $\H2$ is filling the volume relatively uniformly, or $\propto{}l^2$ if the $\H2$ is confined to a disk. However, the scale dependence shown in Fig. \ref{fig:scaleScatter} seems to be much shallower.

In order to understand this better, we now include a log-normal (intrinsic) scatter of the SFRs\footnote{In the remainder of this section we refer to instantaneous SFRs.} in equation  (\ref{eq:SFR}), with $\sigma/\ln{}10$ ranging from 0.1 to 1 dex on the 65 pc scale. Fig.~\ref{fig:scaleScatter2} shows that this scatter decreases with increasing averaging scale roughly as a power law $\sigma_l\propto{}l^{-\alpha}$ with exponent $\alpha\approx{}0.5$.

As we discuss in more detail in Appendix \ref{sect:dep} the rather gradual decline is caused by both the finite width $\sigma_\rho$ of the $\H2$ density distribution and the geometrical arrangement of the molecular hydrogen. In particular, we find that  
a sensible value $\sigma_\rho\approx{}1.5$ \citep{2007ARA&A..45..565M} and a two-dimensional (2D, disk) arrangement of the molecular hydrogen naturally leads to $\alpha\approx{}0.5$.
On the other hand, a purely one-dimensional (1D) arrangement of the $\H2$ is only consistent with $\alpha\approx{}0.5$ if the width in the density distribution is very small ($\ll{}1$), while a three-dimensional (3D) configuration would require $\sigma_\rho\approx{}2.5$. We have also tested that the exponent decreases if the inserted intrinsic scatter becomes large, i.e., $\sigma\gg{}1$. In fact, the top curves ($\sigma=2.3$) in Fig.~\ref{fig:scaleScatter2} clearly show this flattening. Since the $\H2$ density distribution depends on the metallicity and radiation field (see Fig.~\ref{fig:regfit}), $\alpha$ should depend on it, too. With the data points shown in  Fig.~\ref{fig:scaleScatter2}, we obtain $\alpha=0.52\pm{}0.04$ for $Z=0.1$, $\UMW=100$, and $\alpha=0.43\pm{}0.04$ for $Z=1$, $\UMW=0.1$.

Let us now consider the case in which a scatter  $\tilde{\sigma}_l$ is inserted on scale $l$ (we assume a set of discrete scales that change by a factor 2). If the different scatter contributions add in quadrature, the total scatter $\sigma_l$ on a given scale $l$  is simply given by 
\begin{eqnarray}
\sigma^2_l  &=&  \tilde{\sigma}^2_l +  \frac{1}{2^{2\alpha}}\tilde{\sigma}^2_{l/2} + \frac{1}{4^{2\alpha}}\tilde{\sigma}^2_{l/4} + \ldots \\
                      &=&  \tilde{\sigma}^2_l +  \frac{1}{2^{2\alpha}}\sigma^2_{l/2}.
\end{eqnarray}
With knowledge of $\alpha$, this equation allows the computation of the amount of scatter $\tilde{\sigma}_l$ that is introduced on scale $l$ from the measurement of the scatter on scales $l$ and $l/2$. Presumably, different physical mechanisms may introduce different amounts of scatter on different scales. Studying the scale dependence of the scatter may therefore be helpful to uncover the responsible physical mechanism(s). 
%Our simulations also predict that the scatter should never increase by more than $2^{\alpha}$ when averaging scales decrease by a factor of 2.

\section{Discussion}
\label{sect:Discussion}

The scatter in the \MKS{} relation on the scale of $\sim{}100$ pc has been attributed to the evolution of molecular clouds over their life time (see, e.g., \citealt{2010arXiv1009.1971O, 2010arXiv1009.1651S}). In this picture, young molecular clouds have not yet formed stars, but contain large amounts of $\H2$ and hence fall ``below'' the average \MKS{} relation.
On the other hand, clouds that are near the end of their lives are heavily star forming and/or have lost some fraction of their molecular hydrogen, hence they lie ``above'' the relation.
This picture cannot be reconciled with an $\H2$-based star formation law of the form of equation (\ref{eq:SFR}) as long as the gas consumption time scale $\tau_{\rm SFR}/\epsilon_{\rm SFR}$ is treated as a constant. Hence, this explanation of the scatter in the \MKS{} relation implies that $\epsilon_{\rm SFR}/\tau_{\rm SFR}$ has to be a time-dependent quantity. If $\tau_{\rm SFR}$ is approximately constant, then the star formation efficiency will need to change over the life time of a molecular cloud (e.g., \citealt{2010arXiv1007.3270M}, but see  \citealt{FeldmannM}). 

Our interpretation is different. We show that a large amount of scatter in the \MKS{} relation can be explained by the fact that observations do not measure the instantaneous rate of star formation, but rather the number of stars that formed within a finite time interval in the past. Our numerical models predict that the scatter seen on scales of $\sim{}100$ pc  should be small ($\lesssim{}0.1$ dex) \emph{if} SFRs are measured instantaneously. By contrast, if the scatter in the relation is mainly due to an evolving star formation efficiency, the scatter on 100 pc scales should not be strongly diminished if (close to) instantaneous SFRs are used.

We note that the particular small-scale model of star formation used in our simulations is based directly on the $\H2$ density (equation \ref{eq:SFR}). Although this model has been motivated analytically \citep{2009ApJ...699..850K} and is widely used in numerical simulations \citep{2009ApJ...697...55G} or, more recently, semi-analytic models \citep{2010MNRAS.tmp.1359F}, it is important to verify how accurately it describes reality. A potential shortcoming of equation (\ref{eq:SFR}) is that it assumes that a fixed mass fraction of molecular hydrogen is available/eligible for star formation. It has been known for a while now that star formation in molecular clouds occurs preferentially in the region of high gas density ($n\gtrsim{}10^4$ cm$^{-3}$, see e.g., \citealt{1992ApJ...393L..25L, 2004ApJ...606..271G,2010arXiv1009.2985L}). Hence, the SFR should be strongly dependent on the density probability distribution function (pdf) of the gas and not necessarily on the total mass of molecular hydrogen alone.

The density pdf can be expected to depend on details of gas thermodynamics or potential feedback mechanisms (e.g.,  \citealt{2001ApJ...547..172W, 2008ApJ...680.1083R}),
even in a picture in which the turbulence in the ISM is mainly driven by large-scale gravitational motions (e.g., \citealt{2001ApJ...547..172W, 2009ApJ...700..358T}).
Hence, if the SFR is in fact regulated by the amount of high-density gas (and not by $\H2$), we can expect to see differences in the SFR on $\sim$ kpc patches
as a function of $Z$ and $\UMW$, even for the same star formation prescription on small scales. 

In addition, the total amount of $\H2$ may depend on Z and UMW. For instance, the total amount of $\H2$ in the simulated volume changes by a factor of $\sim{}2-3$ when metallicities and radiation fields are varied in the range $Z/Z_\odot=0.1 - 1$ and $\UMW=0.1-100$. This change in the $\H2$ mass alone should induce a galaxy-by-galaxy scatter on the order of $\sim{}0.2$ dex. A study of the scatter of the \MKS{} relation as a function of scale and in regions with different metallicities and UV radiation fields may thus give us, at least in principle, a means to test this picture.

\section{Conclusions}
\label{sect:Conclusion}

\subsection{The scatter in the \MKS{} relation}

Our simulations identify and quantify two important sources of scatter. The first type of scatter is only present if the small-scale star formation relation is non linear and arises due to the scatter in the $\H2$ density at fixed gas density. This scatter is relatively independent of spatial scale and amounts to (at most) $\sim{}0.1$ dex.
The second type of scatter is due to the fact that observations of $\H2$ (or CO), which measure the instantaneous gas surface density, are combined with measurements of SFRs that are averaged over the past tens of megayears. This scatter does not strongly depend on the averaging timescale provided it is longer than both the $\H2$ formation and destruction time, i.e., it exceeds $\sim{}10$ Myr\footnote{Assuming typical conditions $n_{\rm H}\sim{}50$ cm$^{-3}$, $T\sim{}80$ K of the cold neutral medium and $t_\H2=n_{\rm H}/\dot{n}_\H2=1/(R_d\,n_{\rm H})$ with $R_d=6\times{}10^{-18}\,T^{1/2}\,{\rm cm}^{3}\,{\rm s}^{-1}\,n_{\rm H}^2$ \citep{1996ApJ...468..269D}.}. For shorter averaging times, the scatter should decrease and will eventually be dominated by other scatter sources. This type of scatter is strongly dependent on spatial scale. It varies between $\sim{}0.1-0.2$ dex (in $\log_{10}$SFR) on kpc scales and $\sim{}0.4-0.6$ dex on $\sim{}100$ pc scales.

Our numerical experiments predict that intrinsic scatter scales with averaging scale $l$ approximately as $\propto{}l^{-0.5}$. This relatively shallow scaling is primarily caused by the finite width of the $\H2$ density distribution and the arrangement of the molecular hydrogen in a 2-D (disk) configuration.

We note that our simulations provide only a lower limit on the expected scatter as a function of scale, because some sources of scatter (e.g., the velocity dispersion of star clusters and their member stars, or scatter in the $\CO$ to $\H2$ conversion) are not included in our numerical modeling. A precise observational determination of the scatter-scale relation, possibly even as a function of ISM environment, and the comparison with theoretical predictions, such as the one presented in this paper, may thus help to identify the physical processes responsible for creating the scatter. Consequently, we argue that the scale dependence of the scatter in the \MKS{} relation could become an important diagnostic tool in determining the underlying connection between star formation and $\H2$ density. 

\subsection{The environmental variation of the \MKS{} relation}

We have shown that even if the SFR is tightly coupled to the $\H2$ density on small scales (see equation \ref{eq:SFR}), the \MKS{} relation can vary systematically with metallicity and interstellar radiation field in the studied surface density range  $10<\Sigma_\H2 / M_\odot \mathrm{pc}^{-2}<100$, when averaged on 
$\sim{}$ kpc scales.

In particular, the super-linear slope of the \MKS{} relation depends on the actual $\H2$ density distribution and on the existence of a non-linear
scaling between SFR and $\H2$ density. The underlying reason for a slope steeper than unity is that the peak of the $\H2$ density distribution changes systematically with large-scale surface density. At larger $\sim{}$kpc averaged surface densities, more of the molecular gas sits at higher densities, which, due to the non-linear scaling between SFR and density, leads to the super linear steepening of the \MKS{} relation. 

Similarly, the systematic \emph{change} in the slope with metallicity of the ISM and the interstellar radiation field is a reflection of the change in the $\H2$ density distribution. For example, in a low metallicity and/or strong radiation field environment, the HI to $\H2$ transition takes places at significantly higher densities and, consequently, a larger fraction of the $\H2$ mass contributes super-linearly to the SFR. In addition, this implies more star formation at a given $\H2$ surface density and hence changes the intercept of the \MKS{} relation.
  
The scatter in the \MKS{} relation also shows a systematic trend with $Z$ and $\UMW$. The precise value of the scatter and the amount it changes with $Z$ and $\UMW$ depends on (1) the assumed density threshold, $\nc$, above which the SFR scales super-linearly with density, and (2) amount of time over which observed SFRs are time averaged. The scatter varies between $\sim{}$0.05 dex ($Z/Z_\odot=1$, $\UMW=0.1$) and $\sim{}$0.12 dex ($Z/Z_\odot=0.1$, $\UMW=100$) if $\nc=50$ cm$^{-3}$ and SFRs are measured instantaneously. The  trend with $Z$ and $\UMW$
is mainly caused by changes in the $\H2$ and total hydrogen density distributions.
On the other hand, if $\nc$ is large ($\gtrsim{}10^4$ cm$^{-3}$) and the scatter is generated by the time averaging, then the scatter changes only weakly with $Z$ and $\UMW$ ($\sim{}0.09$ dex versus $\sim{}0.13$ dex). 

We note that in order to observe a significant change in the \MKS{} relation, metallicities $\leq{}0.3\,Z_\odot$ and interstellar UV fields $\UMW\geq{}10$ are required. Star-forming galaxies at high redshifts should therefore be the natural candidates to test our predictions.

\acknowledgements
The authors thank A. Leroy and F. Bigiel for stimulating discussions and the anonymous referee for comments that helped improve the manuscript.
The authors are grateful to the Aspen Center for Physics and to the participants and organizers of the workshop "Star Formation in Galaxies: From Recipes to Real Physics". N. Y. G and A. V. K. were supported by the NSF grants AST-0507596 and AST-0708154, and by the Kavli Institute for Cosmological Physics at the University of Chicago through the NSF grant PHY-0551142 and an endowment from the Kavli Foundation. The simulations used in this work have been performed on the Joint Fermilab - KICP Supercomputing Cluster, supported by grants from Fermilab, Kavli Institute for Cosmological Physics, and the University of Chicago. This work made extensive use of the NASA Astrophysics Data System and arXiv.org preprint server.

\appendix

\section{Resolution study}

\begin{figure}
\begin{center}
\begin{tabular}{c}
\includegraphics[width=92mm]{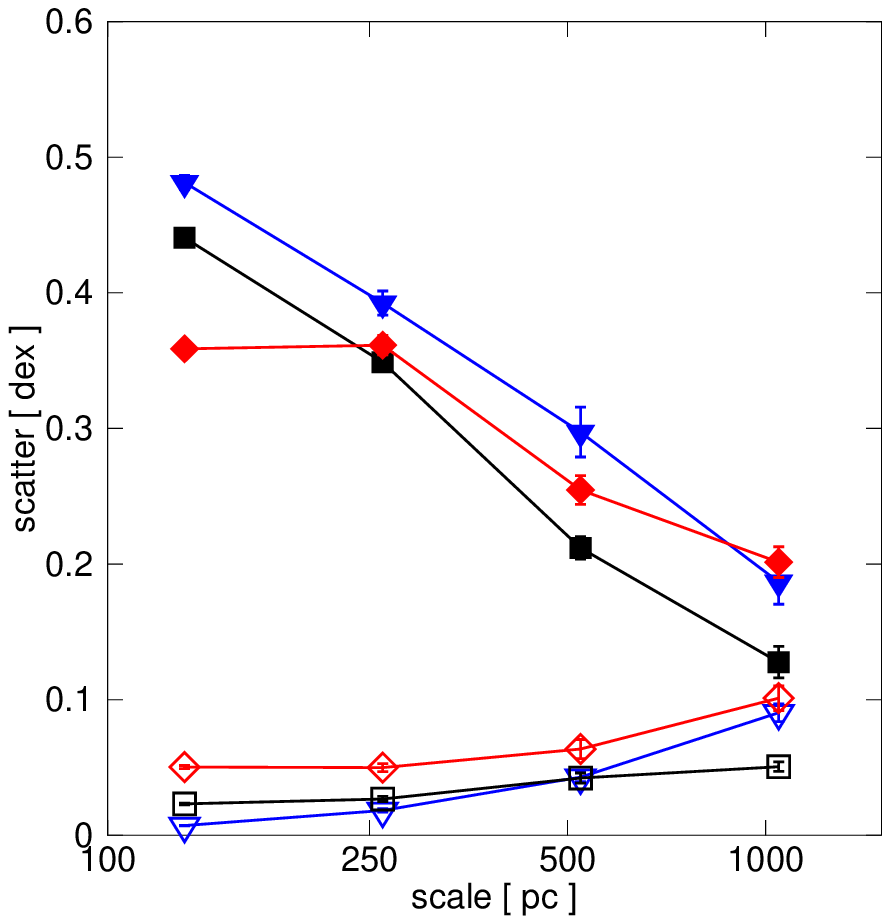} 
%\plotone{f6}
\end{tabular}
\end{center}
\caption{Scale dependence of the scatter of the \MKS{} relation as in Fig.~\ref{fig:scaleScatter}, but now for three different spatial resolutions:
125 pc (blue triangles), 65 pc (black squares), and 32 pc (red diamonds). Empty and filled symbols show the scatter due to a low $\nc$ and SFR averaging, respectively.
\label{fig:resTest}}
\end{figure}

In order to test for possible resolution effects we have rerun one of our simulations ($Z=1$ and $\UMW=1$) at two times better (32 pc) and also two times worse (125 pc) spatial resolution.
We show in Fig.~\ref{fig:resTest} the scale dependence of the scatter in the \MKS{} relation for the three different resolutions (32, 65, 125 pc). As in Fig.~\ref{fig:scaleScatter}, we present both the scatter (1) due to SFR time averaging (using a high $\nc$) and (2) due to the non-linear scaling between SFRs and $\H2$ density (using instantaneous SFR). We find that the amount of scatter as a function of scale is similar in each of the three simulations and that there is no apparent significant systematic trend with resolution.

% figure 6
\section{Dependence of the scatter on spatial averaging scale}
\label{sect:dep}

\begin{figure}
\begin{center}
\begin{tabular}{c}
\includegraphics[width=92mm]{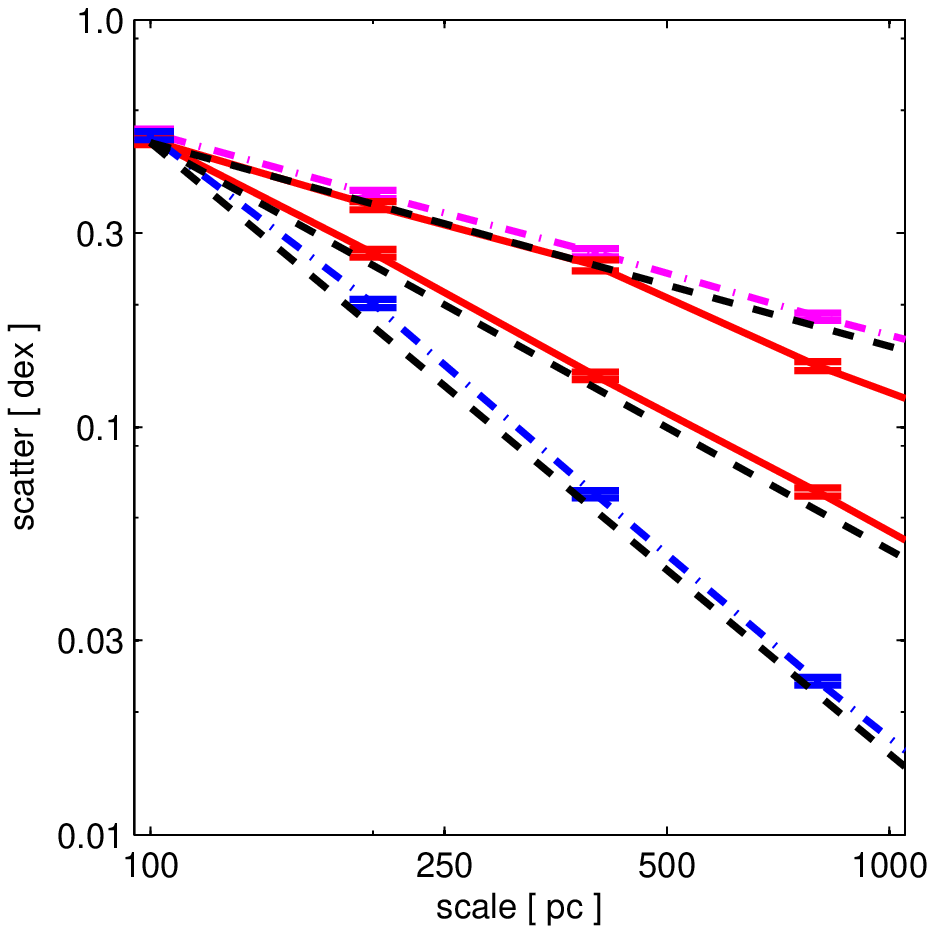} 
%\plotone{f7}
\end{tabular}
\end{center}
\caption{Scatter $\sigma_l$ of the \MKS{} relation as a function of the averaging scale $l$ for different widths of the log-normal density distribution $\sigma_\rho$ and different geometrical arrangements of the molecular hydrogen.
The dot-dashed and solid lines show the result of a simple numerical experiment (see text), in which a spatially uncorrelated scatter with $\sigma{}_{l_{\rm min}}=0.5$ is inserted on the scale of $l_{\rm min}=100$ pc and consequently spatially averaged on larger and larger scales. The dot-dashed lines correspond to  $\sigma_\rho=0.1$
in three (blue, bottom) and one (magenta, top) dimensions, respectively. The red solid lines shows the result for a 2D gas configuration with $\sigma_\rho=0.1$ (lower line) and $\sigma_\rho=1.5$ (upper line). Dashed lines (black) indicate the scaling $\propto{}l^{-0.5}$, $\propto{}l^{-1}$, and $\propto{}l^{-1.5}$ (from top to bottom) and are shown to guide the eye.
\label{fig:scaleScatterTest}}
\end{figure}

As discussed in section \ref{sect:scale}, the scatter in the \MKS{} relation decreases rather slowly ($\propto{}l^{-0.5}$) with averaging scale $l$, much slower than expected from a $V^{-0.5}$ scaling, unless the gas is arranged in a 1D configuration. The scatter $\sigma_l$ in the \MKS{} relation can be formally written as
\begin{equation}
\label{eq:sigSFR}
\sigma^2_l = \langle (\log_{10} \overline{\dot{\rho}_*} - \langle \log_{10} \overline{\dot{\rho}_*} \rangle)^2 \rangle,
\end{equation}
where $\overline{\dot{\rho}_*}=\frac{1}{V}\int{}dx\dot{\rho}_*(x)$ is the spatial average of the star formation rate density 
and the brackets $\langle \ldots \rangle$ denote the ensemble average at \emph{fixed} spatially averaged $\H2$ density $\overline{\rho}=\frac{1}{V}\int{}dx\rho(x)$.

The star formation rate density $\dot{\rho}_*$ is a random field and its value for a given point in space is defined by equation (\ref{eq:SFR}), i.e.,
\begin{equation}
\label{eq:SFRAlt}
\dot{\rho}_* \propto{} \rho \, e^{\sigma{} X},
\end{equation}
where $\rho\equiv{}\rho_\H2$ and we assume that $\dot{\rho}_*$ is proportional to the $\H2$ density, i.e., $\tau_\SFR$ is a constant. Here $X$ is a Gaussian random field with zero mean and unit variance.

From equations (\ref{eq:sigSFR}) and (\ref{eq:SFRAlt}), it follows that $\sigma_l=\sigma/\ln{}10$ on the scatter insertion scale $l_{\rm min}$, i.e., without spatial averaging.
The reason being that, in this case, $\overline{\dot{\rho}_*}=\dot{\rho}_*$, $\overline{\rho}=\rho$,  $\langle \log_{10}\rho \rangle=\log_{10}\rho$ and thus
\begin{eqnarray}
\sigma^2_{l_{\rm min}} &=& \left\langle \left(\log_{10}(\epsilon \rho)+\frac{\sigma{}X}{\ln{}10} -  \langle \log_{10}(\epsilon \rho) \rangle -  \left \langle \frac{\sigma{}X}{\ln{}10} \right\rangle\right)^2 \right\rangle \\
&=&  \left\langle \left(\log_{10}\rho+\frac{\sigma{}X}{\ln{}10} - \langle \log_{10}\rho \rangle \right)^2 \right\rangle \\
&=& \left\langle \left(\frac{\sigma{}X}{\ln{}10}\right)^2 \right\rangle = \left(\frac{\sigma}{\ln 10}\right)^2.
\end{eqnarray}

Equation (\ref{eq:sigSFR}) can be solved if all cells have the same density $\rho_0$
and $\sigma{}\ll{}1$. In this case $\overline{\epsilon \rho\,e^{\sigma{}X}}=\epsilon \rho_0 \overline{e^{\sigma{}X}}$, $\log_{10}\overline{e^{\sigma{}X}}\approx{}\sigma\overline{X}$, and
\begin{eqnarray}
\sigma^2_{l} &=& \langle (\log_{10}(\overline{e^{\sigma{}X}}) -  \langle  \log_{10}(\overline{e^{\sigma{}X}})  \rangle)^2 \rangle \\
&=& \left(\frac{\sigma}{\ln{}10}\right)^{2} \langle \overline{X}^2 \rangle.
\end{eqnarray}
If $X$ is spatially uncorrelated then $\langle X(x) X(y) \rangle=\delta(x-y)$, $\langle \overline{X}^2 \rangle = 1/V$, and $\sigma_{l}\propto{}V^{-0.5}$.

However, the scaling of $\sigma_{l}$ may be different if X is spatially correlated, $\sigma\gg{}1$, or if the density pdf is not sharply peaked.
We can test whether the latter, i.e., the finite width of a log-normal density distribution, provides a quantitative explanation for 
the scaling of $\sigma_{l}$. To this end, we insert spatially uncorrelated scatter with $\sigma{}_{l_{\rm min}}=0.5$ on the scale of $100$ pc and compute $\sigma_{l}$ as a function of averaging scale for log-normal density distributions of various widths $\sigma_\rho$ and for a 1D, 2D and 3D configuration of $\H2$. Specifically, for each scale $l$  we first generate $(l/l_{\rm min})^d$ independent density values ($d$ is the assumed dimensionality of the gas configuration) drawn from a log-normal distribution with width $\sigma_\rho$ and the same number of corresponding star formation rate density values according to equation (\ref{eq:SFRAlt}). We then compute the average $\H2$ density and SFR over the $(l/l_{\rm min})^d$ ``cells''. This pair of spatially averaged $\H2$ density and star formation rate density constitutes a data point in the \MKS{} relation and the scatter of the relation is computed from 1000 data points generated in this way.

Fig. \ref{fig:scaleScatterTest} shows that if $\sigma_\rho$ is sufficiently small ($\ll{}1$), $\sigma_{l}$ scales as $V^{-0.5}$ as anticipated. Supersonic turbulence simulations, however, predict $\sigma_\rho=\ln(1+\beta^2\mathcal{M}^2)^{1/2}$ with $\mathcal{M}\sim{}5-10$ and $\beta\sim{}0.25-0.5$ \citep{2007ARA&A..45..565M}. Thus, $\sigma_\rho{}\approx{}1.5$ is a more reasonable assumption. In this case, $\sigma_{l}$ scales approximately $\propto{}l^{-0.5}$ if the gas is arranged in a disk.

% figure 7

\bibliographystyle{apj}

\clearpage

\end{document}